\documentstyle[11pt,newpasp,twoside,epsf]{article}
\def\edcomment#1{\iffalse\marginpar{\raggedright\sl#1\/}\else\relax\fi}
\reversemarginpar

\begin{document}
\title{Outbursts of Classical Symbiotics: Multi-Wavelength
Observations of the 2000-2001 Outburst of Z Andromedae}

\author{J. L. Sokoloski$^1$, S. J. Kenyon$^1$, A. K. H. Kong$^1$, P. A. Charles$^2$,
C. R. Kaiser$^2$, N. Seymour$^2$, B. R. Espey$^3$, C. D. Keyes$^3$,
S. R. McCandliss$^4$ A. V. Filippenko$^5$, W. Li$^5$,
G. G. Pooley$^6$, C. Brocksopp$^7$, R. P. S. Stone$^8$ } 

\affil{$^1$Harvard-Smithsonian CfA,
$^2$U. Southampton, $^3$STScI, $^4$JHU,
$^5$U. C. Berkeley, $^6$MRAO, Cambridge,
$^7$Liverpool John Moores, $^8$UCO/Lick}

\vspace{0.7cm}
The outbursts of classical symbiotic systems (SS) could be due to
accretion or thermonuclear instabilities, expansion of a white-dwarf
(WD) photosphere in response to a change in accretion rate, or a
combination of these mechanisms.  Since most classical SS are thought
to be burning accreted material on the surface of a WD, they are
closely related to the supersoft X-ray sources, and the causes of the
outbursts may be similar.  Furthermore, the mass loss during outbursts
could play a role in determining whether SS can become Type Ia
supernovae.  We present here the first results from a program of
multi-wavelength monitoring of SS in outburst designed to address the
above issues.

The prototypical SS Z and was observed at radio through X-ray
wavelengths during its 2000-2001 outburst.  On the rise to maximum,
the optical flux moved through two plateaus (observations began during
the first plateau), and the $U$-band flux, which may reflect both
nebular emission plus direct emission from the hot component, has a
different pattern of variation than the $BVR$ fluxes.  On longer time
scales, the $U$-band and far-UV (FUV) fluxes are correlated.
Significant variations may also be present on time scales as short as
days (Fig. 1a), although the 28-minute oscillation that persisted
through a smaller outburst (Sokoloki \& Bildsten 1999, ApJ, 517, 919)
disappeared near optical maximum.

In the radio, our first observation revealed Z And to be fainter
than usual (Seaquist
\& Taylor 1990, ApJ, 349, 313).  The flux therefore dropped very 
early in the outburst, probably indicating that the thermal radio
nebula shrank as the ionizing flux from the WD was blocked by an
optically thick shell.  Z And was not resolved at 5 GHz with the
MERLIN interferometer near the peak of the optical outburst.  Once the
source had brightened again 4 months later, however, spatial structure
was marginally detected by the VLA, at 15 GHz (Fig. 1b).

P-Cygni profiles are apparent in the FUV spectra, and they evolve on a
time scale of weeks (Fig. 1c).  The first $FUSE$ observation shows a
large amount of cool gas (e.g., singly and doubly ionized C, Fe, and
Si), which only partially covers the source of FUV emission.  Multiple
line features evolve from absorption to emission as the outburst
progresses, and there is evidence for hot gas moving at 100s of km/s,
as well as for significant collisional excitation.

X-ray observations were made at three epochs during the outburst.  The
spectra from the first two observations could not be fit with simple
supersoft black-body models.  Emission is detected above 1 keV in
both, and out to 10 keV in the data from XMM (Fig. 1d).  The high
energy emission provides evidence for shock heating of the red giant
wind as the outburst ejecta collide with it.

\begin{figure}
\plotone{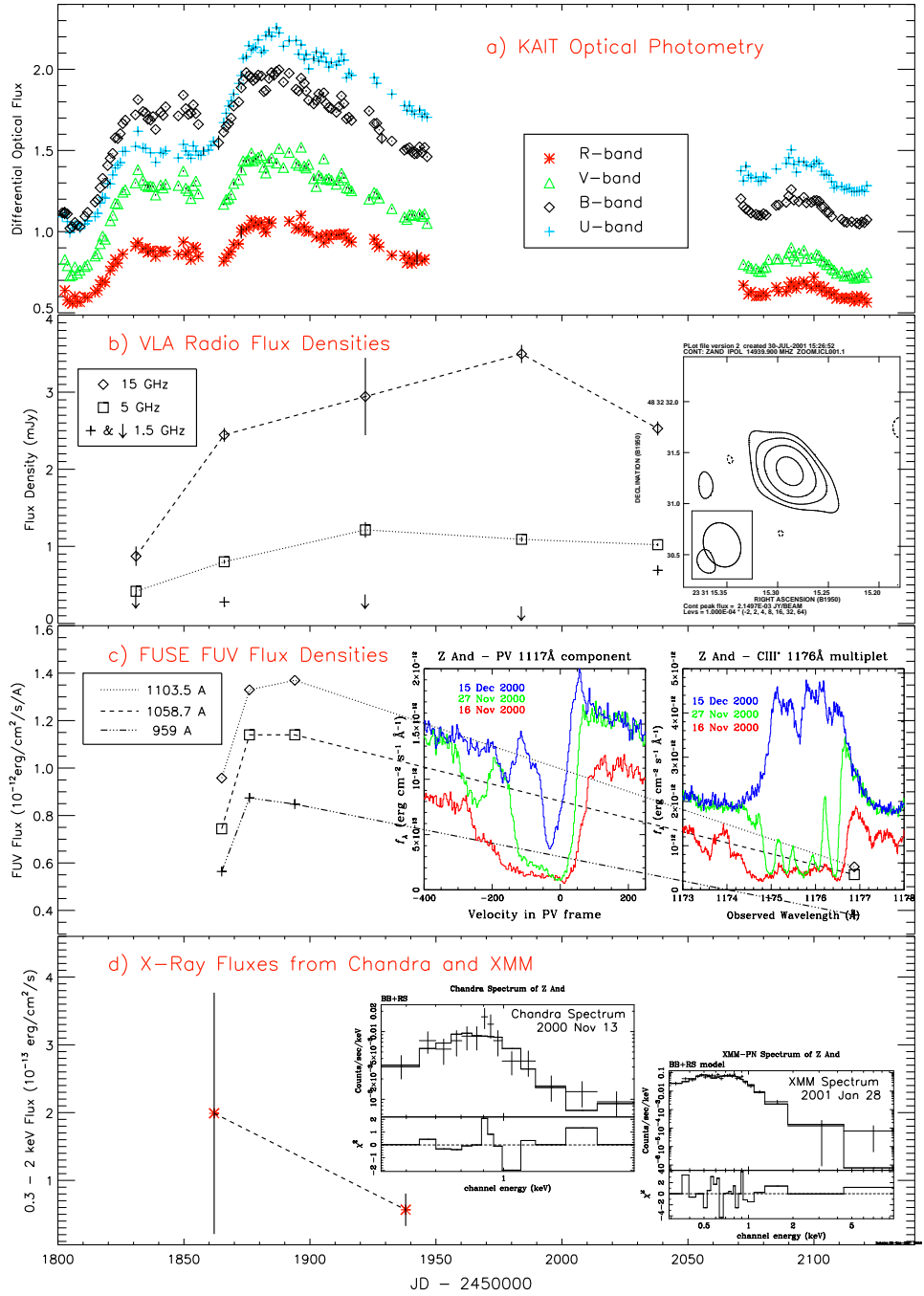}
\caption{\scriptsize a) $U$, $B$, $V$, and $R$ light
curves from the 0.8-m Katzman Automatic Imaging Telescope (KAIT), at
Lick Observatory. 
The light curves have been offset for
clarity. b) VLA radio fluxes at 15 GHz, 5 GHz, and 1.5 GHz.
15 GHz fluxes from the Ryle telescope  
confirm the general trends seen in the VLA data.  In the
2001 May 8 observation, we find some evidence
that the radio extent of Z And 
is resolved at 15 GHz (inset). c) FUV fluxes at three
relatively line-free regions of the continuum. Inset: changes in the
P-Cygni profiles of the P\,V lines, and the evolution from absorption
to emission of the C\,III complex at 1175-1176 \AA. d) Absorbed 0.3 -- 2
keV
X-ray fluxes.
For spectral fits, $N_H$ was set to $5 \times 10^{21}$ cm$^{-2}$
(determined from $FUSE$ data) and abundances were set to solar.  Fit
parameters are $kT_{BB}=81$ (80) eV and $kT_{RS}=0.8$ (0.7) keV for the
Chandra (XMM) data.
}
\end{figure} 


\end{document}